\documentstyle{mn}
\title[Modelling ISO Galaxy Counts]{Modelling ISO Galaxy Counts with Luminosity and Merger Rate Evolution} 
\author[N. Roche and S. A. Eales]{Nathan 
Roche$^{1,2}$, Stephen A. Eales$^{1,3}$\\
$^1$Department of Physics and Astronomy,
      University of Wales Cardiff,
      P.O. Box 913,
      Cardiff CF2 3YB, Wales.\\
{$^2$ \verb"ndr@astro.cf.ac.uk"}\hspace{8mm}   
{$^3$ \verb"sae@astro.cf.ac.uk"}\hspace{8mm}
}

\input{psfig.sty}

\begin{document}

\maketitle
\begin{abstract}
 We model galaxy evolution in the $6.75\mu \rm m$ and 
$15\mu \rm m$ passbands of the ISO satellite, by combining models of galaxy evolution at optical wavelengths (which are consistent with the optical galaxy counts) with observed spectral energy distributions in the infra-red.
Our derived model is consistent with the local $12\mu \rm m$  galaxy luminosity function (from IRAS), including the bright end if we place a small fraction ($\sim 3.5$ per cent at $z=0$) of the model's spiral galaxies  
in interacting pairs with mid-infra-red luminosities enhanced by major starbursts.

Source counts from deep ISO surveys in the $6.75\mu \rm m$ and 
$15\mu \rm m$ passbands significantly exceed non-evolving predictions.
 We find the number counts, redshift distributions and the wide range of mid-infra-red to optical flux ratios of the detected galaxies
 to be reasonably consistent with our evolving model. 
 
Our models suggest that the steep number count of $6.75\mu \rm m$ sources can be explained primarily by the evolving E/S0 galaxies. About one-third of  $6.75\mu \rm m$ sources do appear to be  E/S0s or early-type spirals, consistent with this 
model,  but starburst galaxies at $0.6<z<1.1$ are 
 more prominent in this passband than we expected. 
In the $15\mu \rm m$ passband the main
contributors appear to be evolving, star-forming spirals and starbursting galaxy mergers, as in our model.
 Visibly interacting galaxies form $\sim 31$ per cent of the $15\mu \rm m$
detections to $200 \mu \rm Jy$, and have high mid-infra-red/optical flux ratios 
consistent with our models for major starbursts with high dust extinction 
 (up to $\sim 1.8$ mag in the rest-frame blue-band). The numbers and high mean redshift ($0.68\pm 0.08$)
of merging/interacting galaxies detected in these ISO surveys
  are only consistent with the local mid-infra-red luminosity function if the
merger-triggered starbursts undergo significant luminosity evolution, of  $\sim (1+z)^2$ to $z\sim 1$, in addition to the $\sim (1+z)^2$ number evolution reflecting the increase with redshift in the fraction of merging/interacting galaxies
(as estimated from optical surveys).

\end{abstract}


\section{Introduction}

With the Infrared Space Observatory (ISO, see e.g. Kessler et al. 1996) it has become possible to observe galaxies at mid-infra-red (MIR, $\sim5$--$25\mu \rm m$)
wavelengths to redshifts ($z\sim 1$) at which evolutionary effects become large. The MIR emission from galaxies is closely related to star-formation activity, and well correlated with bolometric luminosity (Spinoglio et al. 1995), and 
much less dependent on the dust content than  the near-ultraviolet emission.
    As the dust extinction in individual star-forming galaxies ranges from near-zero  to  $E(B-V)\simeq 0.9$ or
$\sim 99$ per cent absorption in the near UV (e.g. Calzetti, Kinney and Storchi-Bergman 1994), observing in the MIR reduces a major source of uncertainty in estimating star-formation rates and their evolution.
However, active galactic nuclei (AGN) are also very luminous
in the MIR, and old stellar populations contribute significantly at the shorter MIR wavelengths, so a wide range of galaxies --
not just starbursts  -- may be visible in MIR surveys.

 Three deep surveys have recently been carried out using the ISOCAM instrument
on ISO. The first two covered, respectively, the Hubble Deep Field (HDF) area (Goldschmidt et al. 1997) and a larger field within the Canada-France 
Redshift Survey (CFRS) (Flores et al. 1998a,b), in two  passbands, LW2 ($5\leq \lambda\leq 8\mu \rm m$, centred at $6.75
  \mu \rm m$) and LW3
($12 \leq \lambda\leq 18\mu \rm m$, centred at $15\mu \rm m$), the third 
 a small area in the Lockman Hole field at  $6.75\mu \rm m$ only (Taniguchi et al. 1997).
Most detected MIR sources could be matched with optically detected galaxies or AGN. Spectroscopic data, already available for a high proportion of the
galaxies on the first two survey fields, indicated that ISO was detecting 
a very diverse mixture of extreme starburst galaxies, more normal spirals, ellipticals and Seyferts, out to $z\sim 1$.

The deep ISO surveys found an increase with redshift in the number of galaxies with high MIR luminosities.
Oliver et al. (1997) concluded that their source counts were in excess of 
a non-evolving model at $>3\sigma$, consistent with the  evolving model of
Pearson and Rowan-Robinson (1996) at $15 \mu \rm m$ but steeper in slope than this model at $6.75 \mu \rm m$, and consistent with  the evolving model of 
Franceschini et al. (1994) in both passbands.
However, these models were based on luminosity functions and evolution derived from observations at, respectively, 60 $\mu \rm m$ and optical wavelengths, and may not fully represent the properties of galaxies in the MIR. For example, the $60 \mu \rm m/12 \mu \rm m$ luminosity ratio of galaxies tends to increase with  bolometric luminosity (Rush et al. 1993; Spinoglio et al. 1995),
so the MIR and $60 \mu \rm m$ luminosity functions may differ in shape , and  any class of dust-reddened starburst galaxies will of course be under-represented in optical surveys.

In this paper we compare these ISO surveys with models of our own, considering MIR/optical flux ratios, spectroscopic
and morphological classifications, and redshift distributions of the sources as well as number counts. We aim to improve on previous models by (i)  ensuring that our model's MIR luminosity function at $z=0$ is consistent with the
local luminosity function at $12\mu\rm m$ rather than at
 $60\mu\rm m$, and (ii) including a population of evolving starburst galaxies with properties based on modelled and observed interacting galaxies, in addition to the full range of  normal galaxies  
with  evolution 
consistent with optical observations.
 
Section 2 describes the observational data used in this paper, Section 3 our modelling of evolving spectral energy distributions, k-corrections and luminosity functions for galaxies in the MIR, the derivation of MIR . Section 4 compares our models with the observed number counts, MIR/optical flux ratios, source classifications and redshift distributions. Section 5 discusses the content of the deep ISO surveys, compares published models, and considers future MIR observations.

\section {Observational Data}

Our models of galaxies in the MIR (Section 3.1) are based in part
on the tabulated   spectral energy distributions (SEDs) of Schmitt et al. (1997,
hereafter SKCS).
SKCS combine observations from a large number of sources to give mean low-resolution SEDs for ellipticals, starburst galaxies and Seyferts, from  
radio to X-ray wavelengths. We also make use of (i) the  Mazzei and de Zotti (1994) median ratio of 12, 25, 60 and 100 $\mu \rm m$ fluxes to the $2.2\mu \rm m$ flux for a
larger sample of ellipticals and S0s observed with IRAS,
(ii) the $5.8\leq \lambda\leq 11.5 \mu \rm m$ SED (with additional measurements at $3.6\mu \rm m$ and $12 \mu \rm m$) of the merging starburst 
spiral NGC6090, observed with ISOPHOT on ISO (Acosta-Pulido et al. 1996),
 (iii) the
local $12\mu \rm m$ luminosity functions of non-Seyfert and Seyfert galaxies,  derived by Rush et al. (1993) from an IRAS survey. 

The models are compared with observations from three deep ISO surveys. The first of these, centred on the Hubble Deep Field (HDF),
imaged in the 6.75 $\mu m$ and 15$ \mu m$ passbands (Goldschmidt et al. 1997; Oliver et al. 1997), reaching completeness limits $S(6.75 \mu \rm m)\simeq 40\mu \rm Jy$ and $S(15 \mu \rm m)\simeq 200\mu \rm Jy$, with a few detections at  $S(6.75 \mu \rm m)=15$--$40\mu \rm Jy$ and $S(15 \mu \rm m)=100$--$200\mu \rm Jy$. At  $6.75\mu \rm m$, the survey covered $\sim 6.25$ $\rm arcmin^2$ with resolution $FWHM\sim 4.2$ arcsec, and detected 27 sources at the $3\sigma$ level of which one was identified as a star and 10 as galaxies of known redshift (Mann et al. 1997). At $15 \mu \rm m$, the survey covered $\sim 16.5$  $\rm arcmin^2$ with $FWHM\sim 10$ arcsec resolution, detecting 22 sources of which
10 were identified with galaxies of known redshift. A few redshifts listed by
Mann et al. (1997) are photometric estimates; here these are replaced
by recent spectroscopic redshifts from Aussel et al. (1998).

The second survey covered the larger 100 $\rm arcmin^2$ area of the Canada-France
Redshift Survey (CFRS)
1415+52 field (Flores et al. 1998a, 1998b), reaching $\sim 100\mu \rm Jy$ at $6.75 \mu \rm m$, and a similar limit to the HDF survey at $15 \mu \rm m$.
At $6.75 \mu \rm m$ there were  a total of 59 $3\sigma$
detections of which 7 were identified as stars and 15 as galaxies of known redshift. At $15 \mu \rm m$, the survey detected 78 sources of which two were identified as
stars and  22 as galaxies of known redshift. Part of the CRFS field was imaged
with the HST WFPC2 providing $I$-band morphological data for 16 of the  $15 \mu \rm m$ sources.

In both these surveys the majority of sources without redshifts appeared to 
correspond to 
faint galaxies of unknown redshift, although a significant minority had no visible optical counterparts (Flores et al. 1998a, Aussel et al. 1998).
The third survey (Taniguchi et al. 1997) covered a $3\times 3$ arcmin area within the Lockman Hole field, at
$6.75 \mu \rm m$ only, for an even longer exposure time than the HDF. This survey detected 15 sources to the completeness limit $S(6.75 \mu \rm m)\simeq 32\mu \rm Jy$, but at this point has no spectroscopic follow-up, so for our purposes provides number counts only.  
\section{Modelling Galaxies in the MIR}

\subsection{Spectral Energy Distributions}

We model evolving SEDs for different types of galaxy in the MIR to optical range, by combining the `bc96' models of evolving stellar
populations (e.g. Charlot et al. 1996) with observational data at longer wavelengths. Our models assume $H_0=50$ km $\rm s^{-1}Mpc^{-1}$ and $q_0=0.05$, with galaxies beginning to form stars (with a Salpeter IMF) 16.5 Gyr ago.

Figure 1a shows the mean elliptical galaxy SED from SKCS, replaced  at  $\lambda> 10\mu\rm m$ by the more accurate (much larger sample) Mazzei and de Zotti (1994) SEDs for IRAS ellipticals and S0 galaxies. This is compared with a `bc96' model SED for an elliptical at $z=0$, with an exponentially decreasing star-formation rate (SFR) of short timescale $\tau_{SFR}=0.5$ Gyr, and solar metallicity ($Z=0.02$). The model and observed SEDs are very similar at $0.1< \lambda < 10\mu\rm m$, but at $\lambda\geq 10\mu \rm m$ (log $\nu\leq 13.48$), the bc96 model, considering only direct starlight, predicts no flux, whereas the observed E/S0s
give some emission from infra-red `cirrus',
the general starlight absorbed and re-emitted by small dust grains thinly 
distributed --
with $E(B-V)\sim 0.01$ -- throughout all galaxies (e.g Rowan-Robinson 1992).

Figure 1b shows the SKCS 
mean SED for 11 starburst galaxies of dust extinction $E(B-V)<0.4$, replaced at $3.5\leq \lambda\leq 12 \mu\rm m$ by the ISOPHOT spectrum of the starburst galaxy NGC6090 (Acosta-Pulido et al. 1996), which being of higher resolution 
shows the strong `polycyclic aromatic hydrocarbon'
emission lines in the SEDs of star-forming galaxies at $\lambda=6.2$,7.7,8.6 and
11.6 $\mu\rm m$. These emission lines appear to be characteristic of star-forming galaxies and can have large effects on 
a deep $15\mu \rm m$ survey (Xu et al. 1998).

This SED is compared with a `bc96' model for a galaxy with a constant star-formation rate (SFR), $Z=0.008$ metallicity and
only  $E(B-V)\simeq 0.03$ dust reddening, observed at an age of 1 Gyr.
 The model and observed starburst
SEDs are similar from the $K$-band (2.2 $\mu \rm m$ or log $\nu= 14.14$) to the $4000\rm \AA$ break. In
the near UV the observed SED falls below the model due to greater dust extinction in some of the starburst galaxies; the average extinction  for the sample of 11 is $E(B-V)=0.14$ (from Calzetti et al. 1994). 
At $\lambda>2.2 \mu \rm m$, the observed SED shows a large excess over the model, primarily due to the absorption in dense dust clouds and subsequent reemission in the MIR and far-infra-red (FIR)
of visible and UV light from newly formed massive stars.

Clearly, for
 star-forming galaxies, the `bc96' models are only  appropriate at $\lambda\leq 2.2\mu \rm m$, and an additional component must be added at
longer wavelengths. 
We therefore used the `bc96' models to model the SEDs at  $\lambda\leq 2.2\mu \rm m$, and at longer wavelengths interpolated between the empirical SEDs shown in Figures 1a and 1b; the first representing the MIR/FIR emission from dust heated by an old stellar population, the second representing an actively star-forming galaxy.

\subsubsection{Galaxy SEDs at $\lambda<2.2 \rm \mu m$}
At $\lambda\leq  2.2$ $\mu\rm m$ the bc96 models are used for all galaxies,
with a range of SFR histories for different galaxy types. The SFRs in the elliptical and S0 models decrease exponentially with $\tau_{SFR}=0.5$ and 1.0 Gyr respectively, giving essentially passive evolution which is consistent with the evolution
of E/S0s in optical passbands (e.g. Schade et al. 1996; Roche et al. 1998a). 
In spirals, the SFR is assumed to rise and then fall with time as 
$(t/\tau_{SFR})^{0.5}{\rm exp}(-t/\tau_{SFR})$, with $\tau=4.0$, 5.5 and 8.0 Gyr
for Sab, Sbc and Scd types respectively. The Sab models also contain a 
bulge component; 40 per cent
of their stars are formed as in the S0 model. 

These models predict similar colours and luminosity evolution for the spirals as  the more
complicated spiral evolution models of Roche et al. (1998a) and Roche and Eales (1998). As in these models, we also include a class of  late-type (Sdm) starburst galaxies, with an SED modelled as a constant SFR model at an age of 1 Gyr (giving the SED shown on Figure 1b). Solar ($Z=0.02$) metallicity is assumed for E, S0, Sab and Sbc galaxies,
$Z=0.008$ metallicity for the Scd and Sdm-starburst models. Dust extinction,
modelled as described
by Roche and Eales (1998) with the dependence on wavelength from Mathis (1990),   is added but is small ($<0.1$ mag in the blue-band) for the above galaxy types at the redshifts considered here.

\subsubsection{Galaxy SEDs at $\lambda>2.2 \rm \mu m$}
At $\lambda>2.2 \rm \mu m$ the modelled galaxy SED is an interpolation between the empirical S0 and starburst SEDs. The relative contributions of each are determined by the ratio of the near-UV  ($2800\rm \AA$) and $K$-band 
($2.2 \rm \mu m$) luminosities in the `bc96' model SED, prior to the inclusion
of any dust extinction.

 In the SKCS observed mean SEDs for ellipticals and low-dust starbursts, the $L(2800\rm \AA)/L(2.2)$ ratios are 0.056 and 2.04 respectively. The average dust reddening of the galaxies in the starburst sample,
$E(B-V)=0.14$, corresponds (from the extinction curve
of Mathis 1990) to a 0.90 mag reddening of the $L(2800\rm \AA)/L(2.2)$ ratio, hence before  dust extinction, starburst galaxies would have on average $L(2800\rm \AA)/L(2.2)\simeq 4.68$. For a modelled galaxy SED, a
`UV index', $\alpha_{UV}$, is defined as
as $\alpha_{UV}=(L(2800\rm \AA)/L(2.2)-0.056)/(4.68-0.056)$, giving $\alpha_{UV}=0$ if the $L(2800\rm \AA)/L(2.2)$ ratio before dust extinction is equal to that in the SKCS elliptical SED and $\alpha_{UV}=1$ if it is
equal to that of the SKCS starburst SED.

At $\lambda>2.2$  $\mu\rm m$ the SEDs of all galaxies apart from ellipticals are modelled as $\alpha_{UV}S_{SB}+
(1-\alpha_{UV})S_{E}$, where $S_{E}$ is the SKCS elliptical SED at $\lambda\leq
10 \mu \rm m$ and the Mazzei and de Zotti (1994) S0 SED at longer wavelengths, and $S_{SB}$ the SKCS low-dust starburst SED replaced at $3.5\leq \lambda\leq 12 \mu \rm m$ by the Acosta-Pulido (1996) SED for NCG6090. These two SEDs are normalized
to give continuity with the `bc96' model at $2.2 \mu\rm m$.
 Ellipticals are modelled differently from S0s in that the  Mazzei and de Zotti (1994) elliptical
SED is adopted at $\lambda>10 \mu \rm m$.

This interpolation results in the modelled MIR/FIR luminosity being approximately   
proportional to the $2800\rm \AA$ luminosity, with small offsets depending on 
$L(2.2)$, e.g. $L(15)= 0.88 L(2800 \rm \AA)-0.025 L(2.2)$. The $L(15 \mu\rm m)/L(2800\rm \AA)$ ratio 
is discussed further in Section 5.2.
\subsubsection{`ULIRGs' and AGN}
In addition to the normal galaxies described above, a  significant fraction of FIR and MIR sources are `Ultraluminous Infra-red Galaxies', ULIRGs (e.g. Sanders and Mirabel 1996). The prominence of both ULIRGs and AGN in both local and deeper FIR/MIR surveys required that they be included in our model.
 
ULIRGs are generally found to be gas-rich galaxies in a late stage of merging, 
undergoing intense starbursts with typical ages  $\sim 10$-100 Myr (e.g. Genzel et al. 1998). ULIRGs might be well-described by a 
merger model of Mihos and Hernquist (1996), in which  
two disk-plus-bulge galaxies of similar mass produce a brief
($\sim 50$ Myr duration), intense starburst in the final stages of merging.

On the basis of the observed and modelled properties, we 
model the $\lambda<2.2 \rm \mu m$ SED using a `bc96' constant SFR model, with solar metallicity, seen at an age 0.05 Gyr. The $\lambda>2.2 \rm \mu m$
SED is derived from this as in Section 3.1.2 above.
 We note that real galaxies representing the two types of starburst galaxy in our model -- merging $L\sim L^*$ spirals (ULIRGs), such as NGC6090 (Acosta-Pulido et al. 1996) and low-metallicity dwarf starburst galaxies (e.g. Metcalfe et al. 1996) 
-- have been found to show very similar emission line spectra at  $6\leq \lambda\leq 12 \mu \rm m$. Only in AGN-dominated ULIRGs (Genzel et al. 1998) are the emission lines less pronounced. Hence our model's use of the NGC6090 SED at
 $6\leq \lambda\leq 12 \mu \rm m$ should be a reasonable approximation for star-forming galaxies in general.

Despite their high content of massive stars, nearby
ULIRGs generally have redder colours than normal spirals,
(Sanders et al. 1988). Similarly, the median $I-K$
colour of the detected $15 \mu \rm m$ sources on the CFRS is 0.5 mag redder than  that of the $I$-selected CFRS galaxies (Flores et al. 1998b). These observations indicate that ULIRGs are generally dusty, and on this basis we add high dust extinction of $E(B-V)=0.45$ or 1.8 mag in the blue-band to our merger-starburst model SED, an amount chosen for
consistency with the colours of observed ULIRGs, and also to fit the high MIR/optical ratios of
many ISO sources (Section 4.2).

Finally, our model includes the  Active Galactic Nuclei (AGN). 
Figure 1c shows the SKCS mean SED of a sample of 
15 Seyfert 2 AGN, which are strong emitters in the MIR but without the
prominent emission lines of starbursts. As QSOs, Seyfert 1 and Seyfert 2 galaxies appear to have similar SEDs in the  $2.2<\lambda<30\mu \rm m$ range (Spinoglio et al. 1995),  the SKCS mean SED is adopted for all AGN in the model.
It is assumed that the AGN SED retains the same shape with increasing redshift, but evolves in normalization to give the luminosity evolution described in Section 3.3.
\begin{figure}
\psfig{file=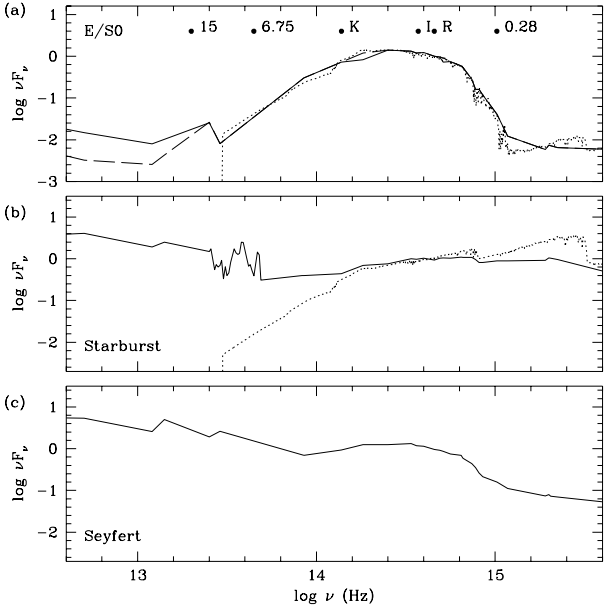,width=85mm}
\caption{(a) The mean SED of a sample of elliptical galaxies (solid line), compared with a `bc96' model for an elliptical (dotted); (b) The mean
SED of starburst galaxies (solid), compared
 with the `bc96' model for 1 Gyr of constant star-formation; (c) The
mean SED of Seyfert 2 galaxies. All SEDs are normalized to log $\nu F_{\nu}=0$
 at $7000\rm \AA$.} 
\end{figure}
\bigskip
 Figure 2 shows the modelled SEDs at $z=0$ of all the types of galaxy in our
model. The merger-starburst SED is shown with and without dust extinction, illustrating that the effects of dust are not significant in the MIR  but  increasingly reduce the flux bluewards of $\lambda\sim 1\mu \rm m$, with the result that our modelled  merger-starbursts 
 have even higher MIR/optical ratios than the Sdm-starbursts,
but relatively red optical colours, similar to Sab spirals.

 Table 1 gives the UV index and three MIR/optical flux ratios for  the galaxy models  at $z=0$. As expected, $\alpha_{UV}\sim 0$ for the E/S0s and $\alpha_{UV}\sim 1$ for the two types of starburst, with
intermediate $\alpha_{UV}$ and  flux ratios for the normal spirals.  
\begin{figure}
\psfig{file=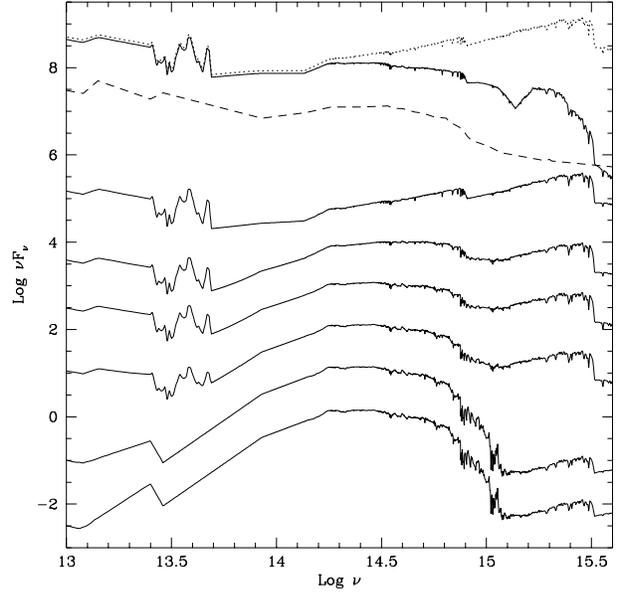,width=85mm}
\caption{Modelled  $z=0$ SEDs of (solid lines, bottom to top) elliptical, S0, Sab, Sbc, Scd, Sdm-starburst and Merger-starburst galaxies, arbitrarily offset on the y-axis. The dotted line shows the merger-starburst SED without dust extinction,  the dashed line the SKCS Seyfert 2 SED.} 
\end{figure}

\begin{table}
\caption{The index $\alpha_{UV}$, and the flux ratios (i)  log $F_{\nu}(6.75 \mu\rm m)/F_{\nu}(I)$, (ii) log $F_{\nu}(15 \mu\rm m)/F_{\nu}(I)$ and (iii)
log $F_{\nu}(12 \mu\rm m)/F_{\nu}(R)$ for our model $z=0$ galaxies and the Seyfert 2 SED from SKCS}
\begin{tabular}{lcccc}
\hline
Galaxy type & $\alpha_{UV}$ & (i) & (ii)
 & (iii) \\
\smallskip
E  & 0.0 & -0.463 &  -0.542 & -0.246 \\           
S0 & 0.0 & -0.465 &  -0.449 & -0.256 \\
Sab & 0.029 & -0.190 & 0.234 & 0.257 \\
Sbc & 0.094 & 0.101 &  0.642 & 0.616 \\
Scd & 0.170 & 0.180 &  0.752 & 0.695 \\
Sdm-starburst & 0.929 & 0.757 & 1.365 & 1.226 \\
Merger-starburst & 1.173 & 1.158 & 1.767 & 1.741 \\
Seyfert & - & 1.042 & 1.610 & 1.569 \\
\hline
\end{tabular}
\end{table}
\subsection{Luminosity Functions}

The other principal ingredient of these models is the galaxy luminosity
function (LF). As the shape of the LF is likely to depend on the wavelength of observation, the most appropriate local reference point may be the $12\mu \rm m$ LFs of low redshift ($z_{mean}=0.013$) galaxies and Seyferts, derived by Rush et al. (1993) from an (almost) all-sky IRAS survey. Figure 3 shows these LFs, converted to $H_0=50$ km $\rm s^{-1}
Mpc^{-1}$, as a dashed line for Seyferts and as the data points for non-Seyfert galaxies.
 
   For the AGN, our model adopts the  
Rush et al. (1993) Seyfert LF, shifted in luminosity according to
the $L(15/L(12)$ and $L(6.75)/L(12)$ ratios from the SKCS Seyfert SED. 
For non-AGN galaxies,  we convert
optical ($R$-band) LFs of the different Hubble types 
to $12\mu \rm m$ LFs, by applying the $10^{40.08}$ correction from characteristic absolute magnitude $M^*_{R}$ to luminosity, with the red-band flux $F(R)=2870\times 10^{-0.4R}$ Jy, and multiplying by our modelled $F_{\nu}(12 \mu\rm m)/F_{\nu}(R)$
ratios (Table 1) and $\nu(12\mu \rm m)$ to obtain a characteristic $12 \mu\rm m$ luminosity
($\nu F_{\nu}$) for each type of galaxy.

As in Roche and Eales (1998), the
red-band galaxy LFs are those derived by Bromley et al. (1998) from the Las Campanas redshift survey, divided into six spectral types, and corrected for incompleteness. These LFs have steeper faint-end slopes for bluer types of galaxy, with $\alpha=-1.89$ for
the bluest (sixth) class. We modelled
 half of the galaxies in the bluest class LF with the 
evolving Scd model (these could represent Sdm/Irr galaxies seen in a non-starburst phase), and grouped these with the spirals on Figure 3, and the other half with the Sdm-starburst SED, and plot these separately.

On Figure 3 we see that, firstly, the steep faint-end slope of the optically bluest galaxies can account very well for the
similarly steep faint end ($\alpha\simeq -1.8$) of the $12\rm \mu m$ galaxy LF.
Secondly,
 at the bright end of the $12\rm \mu m$  LF, the 
E/S0, spiral and Sdm-starburst LFs all cut off at too low a luminosity to explain the numbers of IRAS-detected galaxies with 
$L(12)>10^{41.1}$ ergs $\rm s^{-1}$. 

In order to fit the bright end, we introduce a population of ULIRGs, modelling its LF  by assuming a
fraction $f_{m}$ of Sab, Sbc and Scd galaxies to any any given time be part of a merging/interactioning pair undergoing a starburst. In ULIRGs, the two nuclei are close, with separations from near-zero to $\sim 6 h^{-1}$ kpc (Genzel et al. 1998),
corresponding to $<4.2$ arcsec (the ISO resolution at $6.75\rm \mu m$) at all $z>0.12$, so it is assumed that all ULIRGs are
detected by ISO as single sources. The mass of stars formed in a merger starburst  is less by about an order of magnitude than the mass
of the stars already present in each pre-merger galaxy (from Mihos and Hernquist 1996), but as the starburst 
population is much younger at the time of observation its ratio of optical
luminosity to mass will be about an order of magnitude higher. We therefore  assume the merger starburst to produce the same red-band luminosity as one normal spiral (of which some will be absorbed be absorbed by the greater dust extinction in ULIRG), which seems consistent with the observed  $2.7L^*$ median $R$-band luminosity of ULIRGs (Sanders and Mirabel 1996).

On the basis of this assumption, the $L(12)$
of the merger-starburst is modelled by multiplying $L(R)$ of one normal spiral
by the $F_{\nu}(12)/F_{\nu}(R)$ ratio of our merger-starburst model without dust extinction, log $F_{\nu}(12)/F_{\nu}(R)=1.341$, giving an $L(12)$ which is higher that of the spiral  by factors 12.13, 5.31
and 4.43 for the Sab, Sbc and Scd models respectively.  
The $12\rm \mu m$ LF of the starburst-mergers is then modelled
by assuming a fraction $f_{m}$ of spirals to be reduced in number by a factor 2 but increased in
$L(12)$ by factors of 2 (for the underlying two galaxies in the ULIRG) plus the
modelled starburst/spiral  $L(12)$ ratios, i.e. 14.13 for Sabs, 7.31 for Scds and and 6.43 for the Scds.

 With $f_{m}=0.035$, the combination of the $R$-band galaxy LF
with our modelled $L(12)/L(R)$ ratios fits
the Rush et al. (1993) observed LF reasonably well, at the bright end and over the full range of $12\rm \mu m$ luminosities (Figure 3).
\begin{figure}
\psfig{file=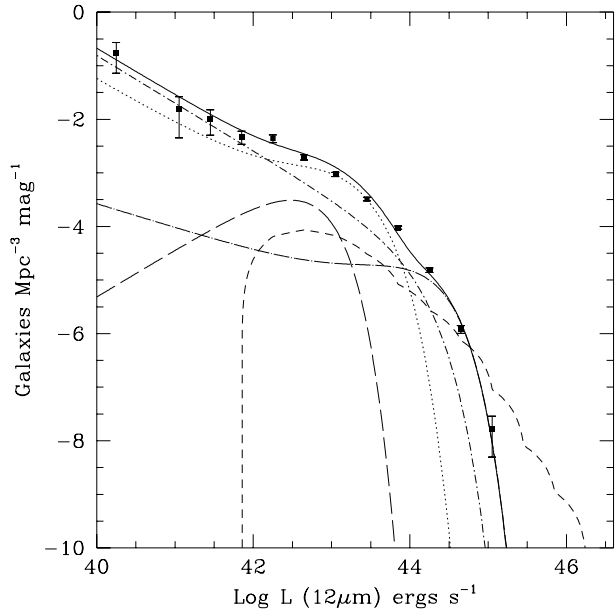,width=85mm}
\caption{Local $12\mu\rm m$ luminosity ($\nu F_{\nu}$) functions of non-Seyfert galaxies (symbols) and Seyferts (dashed line), from Rush et al. (1993); and 
our model's $z=0$ $12\mu\rm m$ luminosity functions
for E/S0 galaxies (long-dashed), spirals and irregulars (dotted), Sdm-starbursts (dot-short dash) and merger-starbursts (dot-long dash), and for all non-AGN
galaxies combined (solid line).} 
\end{figure}

 \subsection{Galaxy Evolution in the MIR}
The $12\mu \rm  m$ LFs derived above are converted to the two ISO passbands using $L(6.75)/L(12)$ and $L(15)/L(12)$ ratios from the SEDs in Figure 2. 
 Non-evolving k-corrections and evolving k+e-corrections, shown of Figure 4,  
are derived by integrating the SEDs from our models over the two ISO passbands
for a range of redshift. 
In the ellipticals and spiral models, the stellar evolution in the `bc96' models produces a corresponding  evolution in the $\lambda>2.2\mu \rm m$ SEDs by 
changing $\alpha_{UV}$ and $L(2.2)$. The 1 Gyr starburst SED is simply redshifted without evolution, so the plotted k+e-correction the same as the  k-correction. The AGN k+e-correction
is the  k-correction combined with luminosity evolution
as $L\propto(1+z)^{3.0}$ at all wavelengths to $z=1.8$, with $L$ remaining constant at $z>1.8$, in agreement with a number of deep AGN
surveys (Page et al. 1998).
 
The merger-starburst galaxies are likely to evolve both in comoving number density and luminosity. The rate of galaxy mergers is predicted (e.g. 
Carlberg et al. 1994) and observed (e.g. Infante, de Mello and  Menanteau 1996;
Roche and Eales 1998) to evolve as $\sim (1+z)^2$, at least at
moderate redshifts. We adopt $f_m=0.035(1+z)^2$ to $z=1.8$, $f_{m}$ constant at  $z>1.8$ (where the luminosity density of AGN and starburst galaxies appears to level out). 
\onecolumn
\begin{figure}
\psfig{file=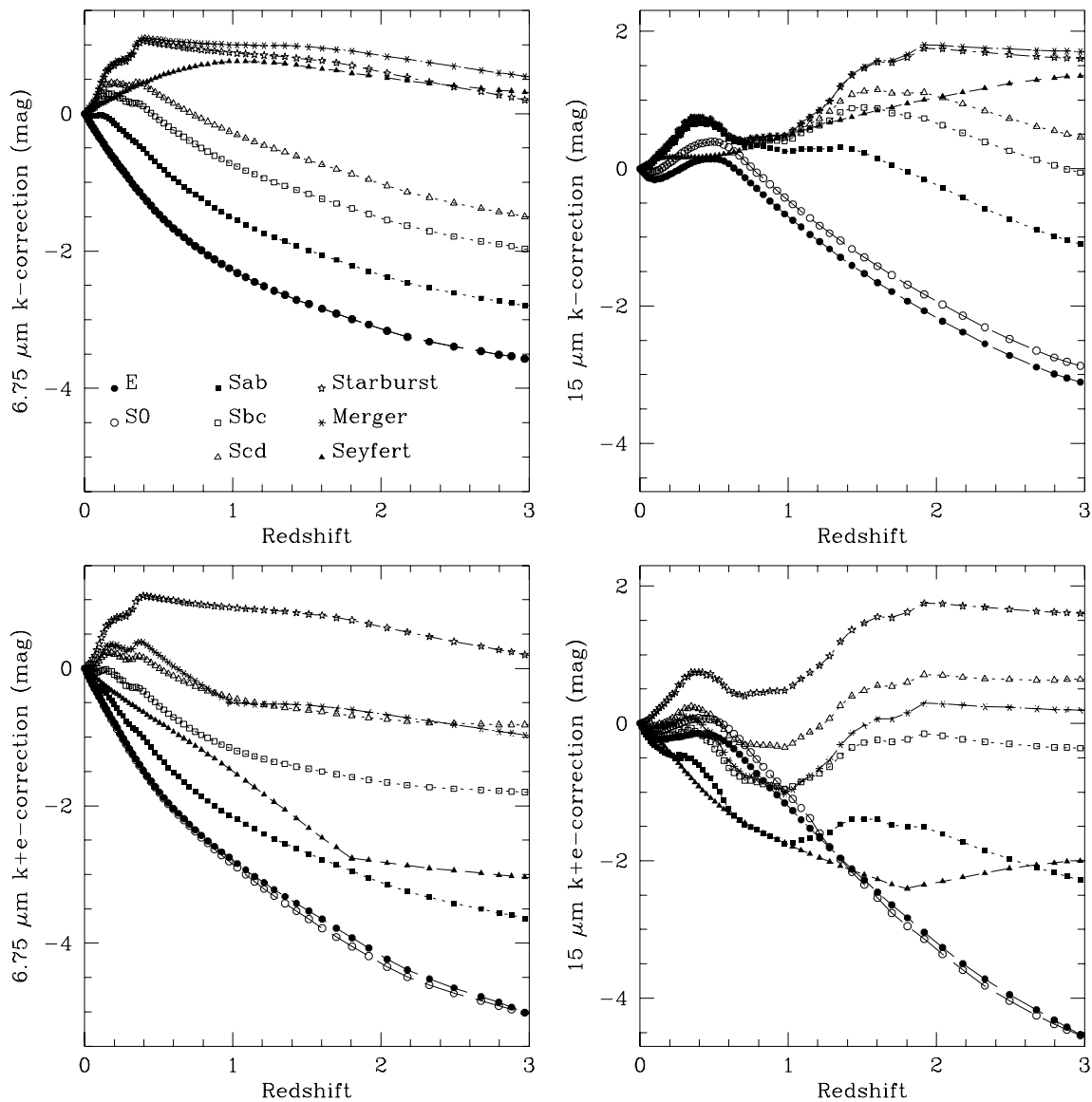,width=170mm}
\caption{Non evolving k-corrections (top) and evolving k+e-correction (bottom)
computed in the $6.75\mu\rm m$ (left) and $15\mu \rm m$ (right) ISO passbands, for our model of seven galaxy types (the ellipticals and S0s are indistinguishable on the upper plots) and for the Seyfert SED of SCKS.}
\end{figure}
\twocolumn

Secondly, the mean luminosity of starbursts in spiral mergers is likely to increase with redshift, due to a higher content of gas at earlier epochs. This increased gas density will be  be reflected in the rate of steady star-formation in normal spirals, which are observed to increase in mean blue-band surface brightness 
with redshift  (Schade et al. 1995; Roche et al. 1998a), approximately as $\sim (1+z)^2$. The luminosity of merger-starbursts might evolve similarly, but at high redshifts may  reach an upper limit --
Weedman et al. (1998) find that starbursts at $2.23\leq z\leq  3.43$ on
 the HDF have on average 4 times the UV surface brightness of their local counterparts. On the basis of these observations, 
we assume $L\propto(1+z)^2$ for  the starburst-merger galaxies  at all wavelengths to $z=1$, with $L$ constant at $z>1$.

On Figure 4, for our models in the in the  $6.75\mu\rm m$ band, the k-correction of early-type galaxies is very negative (i.e. a brightening with redshift), due to the steep rise in the spectrum from the MIR (where luminosity is low due to the absence of star-formation) to the NIR (where there is strong emission from the the old red stars). The bluest galaxies suffer a
k-correction dimming as the observed passband shifts away from the PAH emission features, and for the same reason the evolution of spirals produces only a  small brightening at $6.75\rm \mu m$, so the E/S0 galaxies are expected to become more prominent on going faintwards. 

In the $15\mu \rm m$ passband, the k-corrections of
different galaxy types are more similar, and the effects of the evolution 
of the spirals are much more visible, especially at $0.5<z<1.2$ where the 
PAH emission features are redshifted into this passband. Star-forming galaxies should therefore continue to dominate at faint fluxes.

Our non-evolving model then incorporates the k-corrections on Figure 4, with $f_{m}$ constant, and evolving model uses the k+e-corrections on Figure 4 with $f_{m}$ evolving as described above.

\section{Comparison of Models with Observations}

\subsection{ISO Source Counts}
The differential ISO source counts were obtained (i) for the CFRS field from Flores et al. (1998a, 1998b) by counting the listed $>3\sigma$ detections in 
0.5 mag ($\Delta(\log S)=0.2$) bins, excluding the few  identified as stars, and dividing by the total area 100 $\rm arcmin^2$, and (ii) for the Lockman Hole field from the source list of Taniguchi et al. (1997) with its total area 9  $\rm arcmin^2$, and (iii) for the HDF field from the source lists of Goldschmidt et al. (1997),
including the `supplementary' sources and excluding the one identified star,
and dividing the  count in each bin by the `effective area' plotted by Oliver et al. (1997), which decreases on going faintward to give an approximate correction for
incompleteness. 

Figures 5--8 show these counts with $\surd N$ errors (clustering effects were estimated to increase these statistical errors by only moderate factors of up to 1.3 for the CFRS and 1.15 for the smaller fields). 
The faintest plotted points for the HDF survey, in both passbands, are subject to greater uncertainty than the error bars indicate due to 
large (factors $>5$) corrections for incompleteness and large 
($\sim 40$ per cent) errors on the fluxes of individual sources (which can upwardly bias the observed count if it is  steeply rising). At $6.75\mu \rm m$ the three surveys appear consistent with in the statistical uncertainties,  but in the  $15\mu \rm m$ passband, the CFRS and HDF surveys are 
inconsistent at $10^{-3.4}$ Jy, which might suggest an offset in the flux  
calibrations (Aussel et al. 1998 find some evidence for this). We can only assume the true count lies about mid-way between the two measurements.

\begin{figure}
\psfig{file=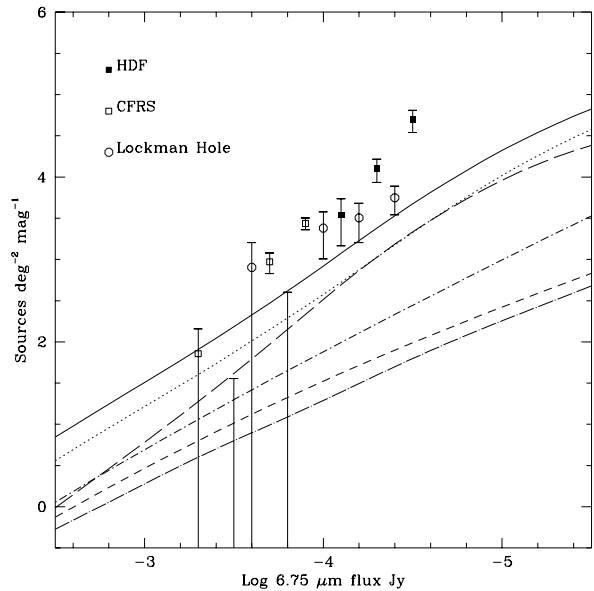,width=85mm}
\caption{Observed ISO cource counts from the CFRS (Flores et al. 1998), HDF (Oliver et al. 1997) and Lockman Hole (Taniguchi et al. 1997) fields, in the $6.75\mu \rm m$ passband, compared with the non-evolving model for E/S0 galaxies (long-dashed),
AGN (short-dashed), spirals/irregulars (dotted), Sdm-starbursts (dot-short-dash) and merger-starbursts (dot-long-dash) and all types combined (solid)}
\end{figure} 
\begin{figure}
\psfig{file=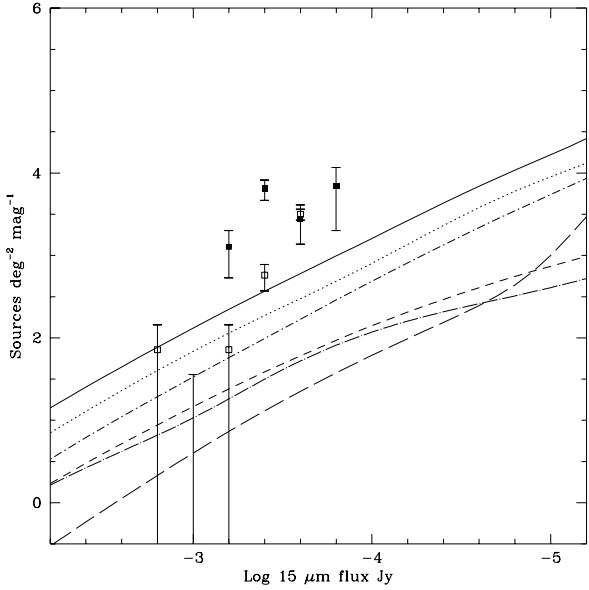,width=85mm}
\caption{As Figure 5 in the $15\mu \rm m$ passband, with CFRS and HDF field
data only.}
\end{figure} 
\begin{figure}
\psfig{file=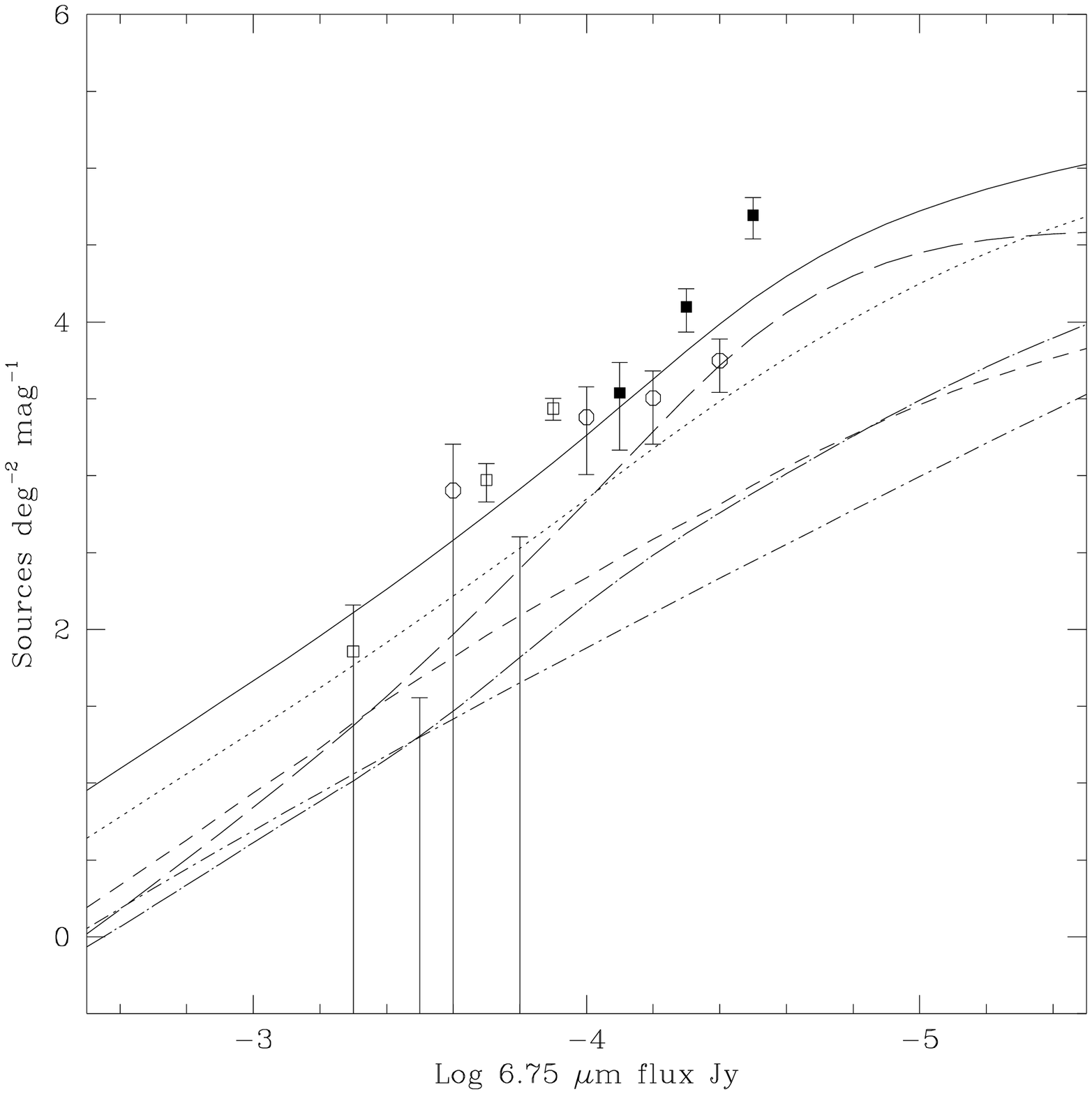,width=85mm}
\caption{As Figure 5, for the evolving model in the $6.75\mu \rm m$ passband}
\end{figure} 
\begin{figure}
\psfig{file=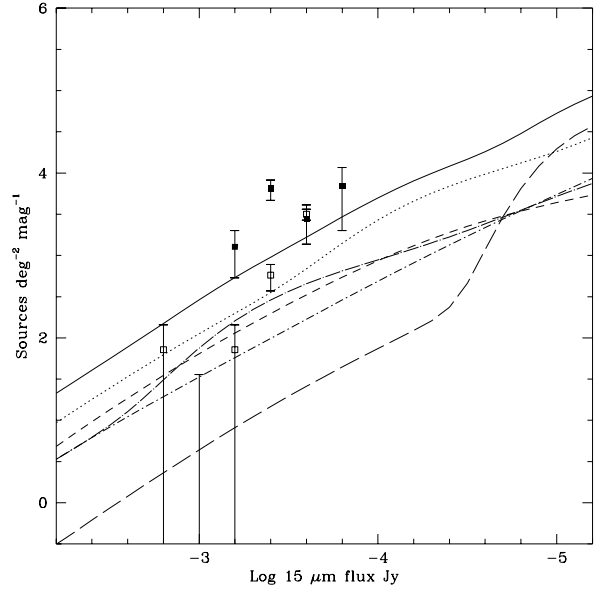,width=85mm}
\caption{As Figure 5, for the evolving model in the $15\mu \rm m$ passband}
\end{figure}

In the non-evolving model in the  $6.75\mu \rm m$ band (Figure 5), the E/S0 galaxies have a steeper number count than other sources due to their very negative k-correction, but this is insufficient to fit the observed count. In
both passbands, non-evolving  models  underpredict the source counts at the fainter fluxes by $\sim 0.5$ dex, a $\geq 3\sigma$ rejection. 

The addition of evolution further steepens the $6.75\mu \rm m$ counts of E/S0
galaxies, to $\gamma\simeq 0.9$ (where $\gamma={d(\log N)\over d(\rm mag)}$).
These evolving E/S0s steepen the predicted count of all sources (at
$\log S\simeq -4.2$) 
from  $\gamma=0.59$ (with no E/S0s) to $\gamma=0.74$, which is consistent with the steep
observed count (Figure 7). At $15\rm \mu m$, evolution increases the  count of spirals and  mergers
in particular (Figure 8), steepening the counts for all source types at
$\log S\simeq -3.7$ from
$\gamma=0.42$ (with no evolution) to $\gamma=0.49$.  In both passbands, the evolving model appears reasonably consistent with the 
the observed source counts, considering the large error bars and the additional uncertainties mentioned above.

Figure 9 shows the $R$-band galaxy number counts predicted using the same evolving and non-evolving galaxy SEDs as in our MIR model (although not including the effect of mergers), compared with
observed galaxy number counts from four CCD surveys.  The non-evolving model underpredicts the galaxy counts by as much as a factor of $\sim 2$, but the evolving model fits well over the whole magnitude range.
 Evidently, the ISO deep galaxy 
counts are at least consistent with the same galaxy evolution as needed to fit the optical galaxy counts, with additional contributions from merger-starbursts and AGN.
However, it is clear that the ISO deep number counts alone give poor constraints on the MIR evolution of galaxies. In the following subsections we make use of the further information provided on the properties of many individual ISO sources in the CFRS and HDF fields. 
\begin{figure}
\psfig{file=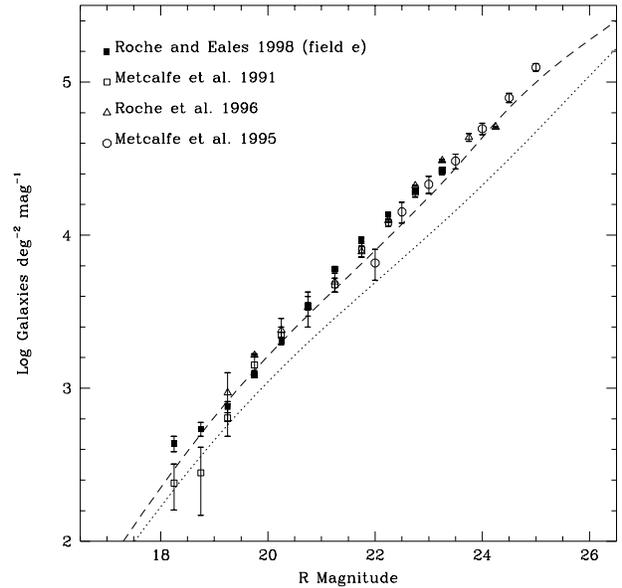,width=85mm}
\caption{The equivalent non-evolving (dotted) and evolving (dashed) models
for the galaxy counts in the $R$-band, compared with observed galaxy counts from
four ground-based $R$-band CCD surveys.} 
\end{figure}

 \subsection{Flux Ratios and Source Classifications}

Flores et al. (1998b) and Goldschmidt et al. (1997)  give
 $I$-band ($\lambda\simeq 0.814 \mu \rm m$) magnitudes for the optical counterparts, in the AB system convertable to flux  as $F(I)=3631\times 10^{-0.4I}$ Jy.
Many of the ISO detections (17 on the HDF area, and 30 on the CFRS field) have identified optical counterparts with redshifts (Mann et al. 1997; Flores et al. 1998a,b; Aussel et al. 1998),
so their flux ratios $F(6.75)/F(0.81)$ and $F(15)/F(0.81)$ as a function of redshift, together with the  spectral and/or morphological classifications of the survey authors,
can give an indication of the range of galaxy
types represented.
 
Figures 10 and 11 show, respectively, observer-frame $F(6.75)/F(0.81)$ and $F(15)/F(0.81)$
ratios of the ISO sources with redshifts, compared to our evolving galaxy models. The  distribution of flux ratios suggests that the range of galaxy types represented in the ISO surveys is comparable to the full range in our model.
\begin{figure}
\psfig{file=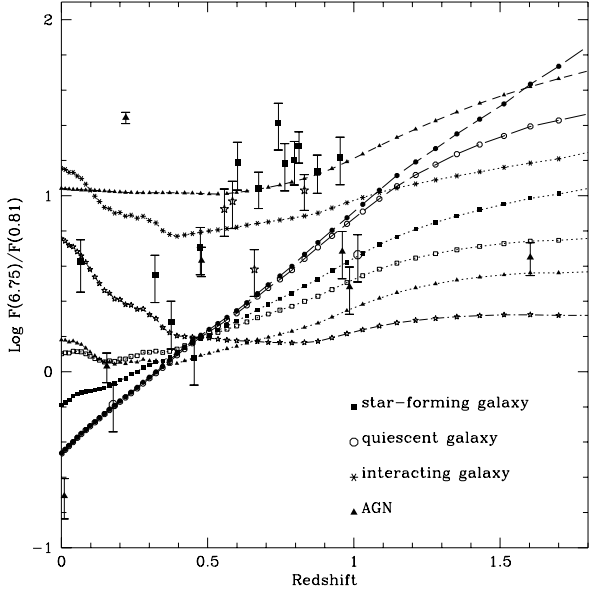,width=85mm}
\caption{The ratio of observer-frame $6.75\mu \rm m$ and $I$-band ($0.81 \mu \rm m$) flux ($F_{\nu}$) for the 24 (CFRS and HDF) $6.75\mu \rm m$ sources  with
redshifts, compared with our models for the different galaxy types, plotted with  the same symbols as on Figure 4.} 
\end{figure}

    Many $6.75 \mu \rm m$ sources do seem to be `normal' galaxies; 9/26 of those
with redshifts have $F(6.75)/F(0.81)$ ratios consistent with our E/S0 or spiral
models, and two more with the Sdm-starburst model, and of 10 of sources 
on the HDF, 4 appeared to be normal spirals and 2 to be ellipticals. However, 14/26 sources have high $F(6.75)/F(0.81)$ ratios
consistent only with the AGN or the 
merger-starburst models, a $\sim 3\sigma$ excess over the 23 per cent fraction of these types in our evolving model at 
 $S(6.75\mu \rm m)\geq 100\mu \rm Jy$. A total of 7/24 have optical spectra indicating the presence of AGN, slightly more than but consistent with the AGN fraction in the model
(15 per cent at $100\mu \rm Jy$). Three out of the 10 $6.75 \mu \rm m$ sources on the HDF are visibly interacting, and also
 have  $F(6.75)/F(0.81)$ ratios consistent with the merger-starburst model.

\begin{figure}
\psfig{file=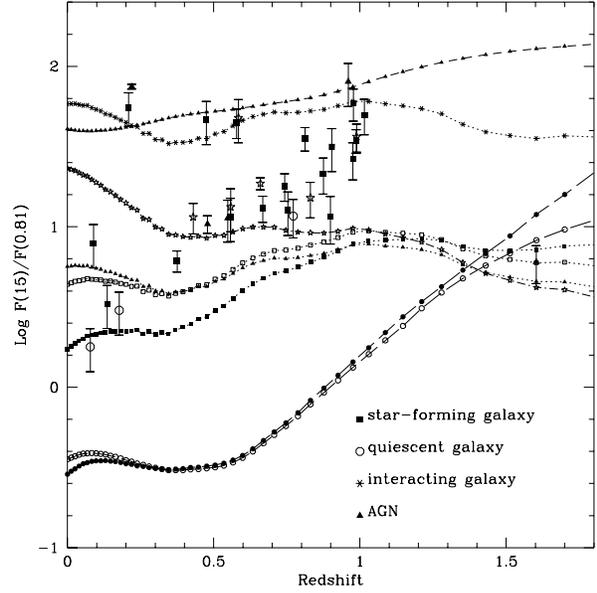,width=85mm}
\caption{The ratio of observer-frame $15\mu \rm m$ and $I$-band ($0.81 \mu \rm m$) flux ($F_{\nu}$) for the 32 (CFRS and HDF) $15\mu \rm m$ sources  with
redshifts, compared with our models for different galaxy types, plotted with the same symbols as on Figure 4.} 
\end{figure}
The $15\mu \rm m$ sample is even more dominated by
starburst galaxies.
Of  32 sources with redshifts, only one has a  $F(15)/F(0.81)$
ratio consistent with a non-star-forming galaxy -- and it is a spectroscopically
identified QSO. A further 5/32 have flux ratios consistent with normal spirals at $z<0.4$ , and  8/32 with being either  spirals or   Sdm-starburst galaxies at $0.4<z<1.0$. This seems consistent with our evolving model in which only 1.5 per cent of $S(15\mu \rm m)\geq 100\mu Jy$ sources are E/S0s, but 41 per cent are  spirals and 11 per cent  are Sdm-starbursts. 
 The majority of the
$15 \mu \rm m$ sources, including all five of the known interacting galaxies, have high flux ratios covering the range from the  Sdm-starburst to the merger-starburst
and AGN models, suggesting a diverse mixture of  rapidly star-forming galaxies with a
wide range of  dust extinctions. 

 Flores et al. (1998b) find
6 of the 16 CFRS field sources observed with the HST to be interacting galaxies, as are 2/10 of the  $15 \mu m$ sources on the HDF, giving a total of a $31\pm 11$ per cent
fraction of starburst-merger galaxies, consistent with our
evolving model in which starburst-mergers make up 26 per cent of
 $S(15\mu \rm m)\geq 200\mu Jy$ sources. If the
number and LF of starburst-merger galaxies did not evolve but remained at that needed to fit the local $12\mu\rm m$ LF (Section 3.2; Fig 3), their numbers at this
limit would be an order of magnitude lower (49  $\rm deg^{-2}$ compared to 507 $\rm deg^{-2}$), and  with $(1+z)^2$ density evolution but no
luminosity evolution, they would be a factor of four lower (131 $\rm deg^{-2}$). 

Lastly, the 4/32  fraction of $15 \mu \rm m$ sources identified on the basis of optical spectra as Seyferts or QSOs is consistent with the evolving model
prediction of 21 per cent.

In summary, the evolving model, in addition to fitting the ISO source counts,
is consistent with observed fractions of different source types, with the
exception that there appear to be about twice as many starburst galaxies in the $6.75\mu \rm m$ surveys as predicted. In the $15 \mu \rm m$
passband,
the observed $\sim 31$ percent fraction of starbursting merger galaxies is consistent with our
evolving model, and 
 greatly exceeds that expected with no evolution.

\subsection{Source Redshift Distributions} 
 Finally, the  redshift distributions of the sources can be compared with the models, although with relatively poor statisics. Figure 12 shows the observed 
$N(z)$ for the $6.75\mu \rm m$ sources with redshifts (with the CFRS and HDF
$N(z)$ shown separately as their flux ranges are separate) and the
$15\mu \rm m$ sources (with the CFRS and HDF results combined as the flux limits are similar). We assume that the spectroscopically identified sources are an unbiased sample of all detected sources, and so correct for the incompleteness
of the spectroscopy by multiplying each $N(z)$ by the ratio of the total number of (non-stellar) detections to the number with redshifts. The observations are
compared with  the non-evolving and evolving models. For the $15\mu \rm m$ survey, the models are also shown for the starburst-merger galaxies separately  -- with evolution they make up the majority of the detections
at  $0.6<z<1.1$.

\begin{figure}
\psfig{file=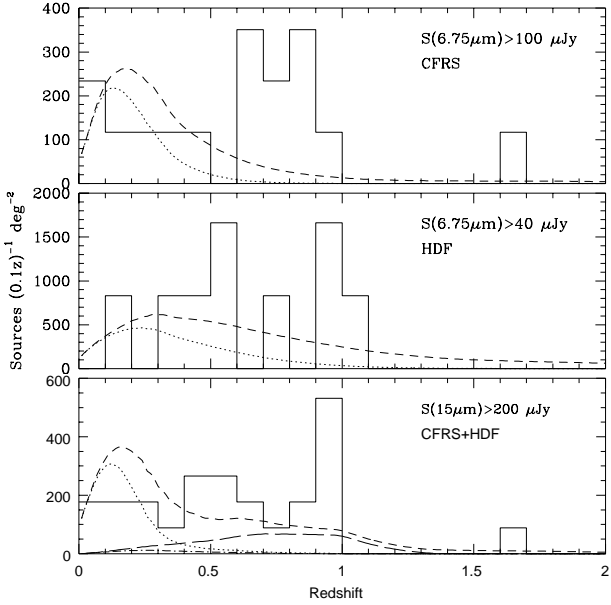,width=85mm}
\caption{The observed redshift distributions of ISO sources (a) on the CFRS field with $6.75 \mu \rm m$ fluxes above $100\mu \rm Jy$ (b) on the HDF 
$6.75 \mu \rm m$ fluxes above $40\mu \rm Jy$ (c) on the two surveys combined with $15 \mu \rm m$ fluxes above $200\mu \rm Jy$; all normalized to the total number of non-stellar detections. These are compared with non-evolving (dotted) and evolving (models) for all galaxy types. On (c), the model $N(z)$ for 
merger-starbursts only is shown with (long-dashed) and without (dot-dash)
evolution.} 
\end{figure}

In the $6.75\mu \rm m$ band, the observed $N(z)$ (with $z_{mean}=0.62\pm 0.10$
and $0.64\pm 0.10$ for $100\mu \rm Jy$  and $40\mu \rm Jy$ limits) is much more extended than the non-evolving model (giving $z_{mean}=0.21$ and 0.38 respectively) and more
consistent with the evolving models ($z_{mean}=0.42$ and 0.78). There may be some excess at $0.6<z<1.1$ over even the evolving model, apparently (Figure 10) made up of star-forming galaxies with high $F(6.75)/F(0.81)$ ratios at least equal to the starburst-merger model (see Section 5.2).

In the $15\mu \rm m$ band, the observed mean redshift of $0.62\pm 0.07$ at
a $200\mu \rm Jy$ limit is much more consistent (by $\sim 4 \sigma$) with the 
evolving ($z_{mean}=0.49$) than the non-evolving model ($z_{mean}=0.18$),
and the shape of $N(z)$ at $0.4<z<1.0$ is similar to the flat extended
`tail' 
of the evolving model -- presumably this is the result of the contribution from PAH  emission lines over this redshift range.

The six identified merging/interacting galaxies  at this limit have $z_{mean}=0.68\pm 0.08$, consistent with the $z_{mean}=0.72$ predicted by our model for merger-starburst galaxies undrgoing both
density and luminosity evolution. Models with the same $z=0$ LF with no evolution and with density evolution only, give, respectively,
$z_{mean}=0.35$ and $z_{mean}=0.46$,  
underpredicting
the mean redshift of merging/interacting galaxies as well as the numbers detected.

Finally, in both passbands there are a few low-redshift ($z<0.3$) sources, in numbers 
consistent with the steep faint-end slope of the model LF.

\section{Discussion}

\subsection{The Content of the Deep ISO Surveys}
We have combined observed infra-red SEDs of star-forming and passive galaxies
 with galaxy evolution models, which also fit the galaxy number counts in the optical passbands, to predict the luminosities and evolution of the different 
galaxy types in the MIR. The model, applied to the local $R$-band LF from the 
Las Campanas survey, appears consistent with the  
local galaxy $12\mu \rm m$ LF, although  to fit the bright end it is
necessary for a small proportion of galaxies to have 
enhanced MIR/optical luminosity ratios -- this was represented in our model 
 by placing 3.5 per cent of spirals into merging or interacting pairs with tidally-triggered starbursts of $R$-band luminosity (before dust extinction) equal to that of each
pre-merger galaxy. 
The steep faint end ($\alpha\sim -1.8$) of the Rush et al. (1993) LF
corresponded to that of the bluest type of galaxy in optical surveys, suggesting that these late-type dwarf galaxies will make up most of the lower redshift MIR sources. 

The deep ISO counts significantly ($\sim3\sigma$) exceed
non-evolving predictions, as previously concluded by Oliver et al. (1997), and
are much more consistent with our evolving model.  Secondly,
the redshift distributions of indentified  ISO sources are much more extended than the non-evolving models, indicating that their high number counts result primarily from
luminosity evolution, as in our model, rather than density evolution. 
 We interpret the success of this evolving model in the MIR as indicating  that deep MIR and optical surveys trace the same star-formation history -- primarily,  that of the various classes of spiral. However, the MIR surveys also show contributions from other galaxy types:

(i) In our model the steepness of the $6.75\mu \rm m$ source counts is accounted for by a significant contribution from E/S0 galaxies (e.g. 29 per cent of sources to $100\mu\rm Jy$; 44 per cent to $40\mu\rm Jy$), which with a combination of
 a strong k-correction brightening with passive evolution consistent with the evolution of ellipticals in the optical passbands (e.g. Schade et al. 1996; Roche et al. 1998a), show 
a very steep ($\gamma\simeq 0.9$) number count.
 The distribution of $F(6.75)/F(0.81)$ ratios (Fig 10) is consistent with $\sim 1/3$ of the  $6.75\mu \rm m$ sources being E/S0s and therefore supports this
interpretation of the steep counts.

(ii) Secondly,  the $\sim 31$ per cent of the $15\mu \rm m$ sources to $200 \mu \rm Jy$ which were visibly interacting pairs had high MIR/optical flux ratios which we interpret as indicating major starbursts with significant dust 
extinction of up to $\sim 1.8$ mag in the rest-frame blue.

 There is already evidence from optical and 
$K$-band observations that the fraction of 
merging and interacting galaxies increases with redshift, approximately as $(1+z)^2$, at least at moderate redshifts (e.g. Infante et al. 1995; Roche and Eales 1998; Roche et al. 1998b).  At $z=0$, the number density and MIR luminosity of the  
starburst-merger population are constrained  by the bright end of the Rush et al. (1993) LF. When compared with a model which fit this local LF, 
the  numbers and  relatively high mean redshift 
($0.68\pm 0.08$) of starburst-mergers in the deep MIR surveys required luminosity evolution in addition to the $(1+z)^2$ increase in comoving number density 
expected from the evolving merger rate. More deep MIR data will be needed to accurately determine this rate of luminosity evolution,  but the observations appeared well-fitted by $L\propto(1+z)^2$ to $z=1$, similar to the surface brightness evolution of
normal spirals, as expected if the change with redshift in the typical luminosity of the merger-triggered starbursts traces the evolving gas content of the pre-merger spirals. 

(iii) Thirdly, some $\sim 20$ per cent of ISO sources are Seyferts or QSOs, as expected from the
already well-determined LF (Rush et al. 1993) and evolution (e.g. Page et al.
1998) of this type of source.
\subsection{Comparison of MIR Models}
We believe that the evolving model considered here, although based on limited 
observational data and many approxmations, improves on previous models of galaxy evolution in the MIR in its consistency with both deep optical surveys and the 
$12\mu \rm m$  LF of local galaxies. Oliver et al. (1997) compare their counts
with two models, one from Pearson and Rowan-Robinson (PRR, 1997) and one from
 Franceschini et al. (AF, 1994). We discuss these two models in turn.

The PRR model, firstly,  contains only two galaxy types (plus AGN), `cirrus galaxies'
(spirals), which undergo little evolution, and starburst galaxies, which evolve 
as $L\propto(1+z)^{3.1}$. However, we find that evolving E/S0 galaxies are important in accounting for the 
steep slope of the $6.75\mu \rm m$ counts. E/S0s are not represented in the PRR model, which consequently predicts too shallow a count slope in the $6.75\mu \rm m$ passband, and was rejected by Oliver et al. (1997).

Secondly, the relations between SFR and luminosity in the PRR model correspond to $L(15)\simeq 0.57L(2800\rm \AA)$, whereas in our model based on the SKCS mean SEDs, $L(15)\simeq 0.88L(2800\rm \AA)$. The difference appears to be related to 
the $L(15)/L(60)$ ratio which is 0.28 in PRR and 0.45 in our models.
IRAS observations (Spinoglio et al. 1995) indicate a typical flux ratio of
$\log(F_{60}/F_{12})=1.1$ for normal star-forming galaxies, which corresponds to
$F_{\nu}\propto\nu^{-1.57}$ and $L(15)/L(60)\simeq 0.45$, as in our model.
The discrepancy between the high Rowan-Robinson et al. (1997) estimates of the SFRs in individual ISO galaxies, and the results of Flores et al. (1998b)
and the global SFRs from UV and FIR surveys, also suggest they are using a model with too little of the total flux in the MIR range.

Thirdly, PRR model the MIR galaxy LF using a shifted $60\mu\rm m$ LF, but as the
 $L(60)/L(12)$ ratio of galaxies tends to increase with bolometric luminosity (Rush et al. 1993; Spinoglio et al. 1995),
the true galaxy LF in the MIR will be steeper than
at $60 \mu m$. In particular, PRR use a faint-end slope $\alpha=-1.27$ for the starburst galaxies, whereas MIR and optical observations give
$\alpha\simeq -1.8$.

Fourthly, in the PRR model most of the evolution is concentrated into the starburst galaxies, which  number about one-seventh of the total, whereas in our model the luminosity evolution is distributed more evenly amongst E/S0s, spirals and merger-starbursts, allowing the counts to be fitted with a more moderate rate of evolution. The latter seems more consistent with optical evidence that E/S0s, early-type and late-type spirals all have similar rates of luminosity and
surface brightness evolution out to $z\sim 1$ (e.g. Schade et al. 1995, 1996; Roche et al. 1998a; Glazebrook et al. 1998). 
Elbaz et al. (1998) claim that a steepening of the counts at $S(15\mu\rm m)\sim 2$ mJy supports an even more extreme division into non-evolving galaxies and  
starburst galaxies evolving almost as $(1+z)^6$. However, the reported steepening   occurs at
exactly the flux where there is a slight steepening of the counts in our evolving model as redshifted emission lines begin to enter this passband, and is based on a single field where statistical errors and/or clustering might have exaggerated the change in slope.

The AF model includes E/S0 galaxies with very similar SEDs and 
evolution as our model, and so does fit the steep $6.75\mu \rm m$ counts. The AF model is much more similar to ours in that it includes the full range of Hubble types, with evolution derived from evolving SFRs, and it appears more consistent with the deep 
NIR and optical galaxy counts than PRR (see the AF and PRR papers). The AF model includes a  starburst population with only slight evolution and a steep LF, similar to
our model's Sdm-starbursts, and predicts the ratio of starburst to normal spiral galaxies to decrease on going faintwards, with only an 18
per cent fraction of starburst galaxies at  $S(15\mu\rm m)= 250\mu \rm Jy$. However, observations suggest that 
the ratio of peculiar to spiral morphologies in ISO surveys increases with redshift (Elbaz et al. 1998), and the high $F(15)/F(0.81)$ ratios (Fig 11) of many sources indicate a starburst galaxy fraction closer to 50 per cent.  
These discrepancies can be resolved simply by adding, as we have done, a second population of
starburst galaxies -- spirals undergoing major mergers with a more rapidly
evolving LF.

 The inclusion of density evolution, corresponding to a $(1+z)^2$ merger rate,  reduces the luminosity evolution required for the merger-starbursts
 from that of AGN or greater (e.g. Elbaz et al. 1998) to a  rate similar
to their spiral progenitors, which is surely more easily explained physically. Luminosity evolution will have a greater effect on mean redshift, relative to its effect on the number counts,  than density evolution; hence our model's  consistency with the mean redshift of 6 interacting 
galaxies (Section 4.3) supports the adopted balance of these two processes, but clearly a much larger sample is needed to confirm this.

At $6.75\rm \mu m$ our model may underpredict the number and  $F(6.75)/F(0.81)$ ratio (Fig 12) of starburst galaxies at 
$0.5\leq z\leq 1.0$. This discrepancy may be the result of our simple
 interpolation of the starburst SEDs in the $3.6\leq \lambda\leq 5.8\mu \rm m$ range where there is
little data available. The SED of the nearby Circinus galaxy (Moorwood et al. 1996), although noisy in this range, suggests the presence of emission features
at $\lambda\sim 4.0\mu \rm m$, which would add to the 
$6.75\rm \mu m$ fluxes from starburst galaxies at  these redshifts --
another possibility is that many of these starburst galaxies have an additional contribution to the flux at this wavelength from obscured AGN. 
\subsection{Future Prospects}
Considering future MIR observations, surveys at  $\sim 15\mu \rm m$, on account of the relatively `clean' sensitivity to star-formation, should be particularly useful for studying various types of starburst galaxy, including the merging spirals and Sdm galaxies of our model, comparing their properties such as dust extinction and determining their 
evolution  to $z\sim 1$, beyond which k-corrections become unfavourable. This redshift has already been reached at $200\mu \rm Jy$. However, to interpret the results it is important that sources are carefully classified  by spectroscopy, high-resolution imaging etc., to study  the rate and type of evolution of each type of galaxy. Hence at this wavelength it is probably most useful for future surveys to be aimed at acquiring much larger source samples to similar
$\sim 0.2\rm mJy$ limits; this should be achieved by the forthcoming WIRE
Deep Survey.

The $6.75\rm \mu m $ passband may be less useful until galaxy SEDs at
$\sim 3$--$6\rm \mu m$ are better determined, thereafter it may be particularly
suited to studying the evolution of  E/S0s and disk-plus-bulge spirals. As  
these types are expected to evolve most rapidly at $z>1$, and as their 
$6.75\rm \mu m$ k-correction becomes increasingly favourable to $z\sim 4$, 
it might be most useful in this passband to survey  small fields to the greatest
possible depths (i.e. the confusion limit) in the hope of reaching the 
  `Lyman break' galaxies at $2.5<z<4.0$.
At these redshifts,
$6.75\rm \mu m $ corresponds approximately to the rest-frame $K$-band, and fluxes will aid our understanding of these objects by giving an indication of  
the mass of stars already formed, rather than the star-formation rate. 

Our model, which assumes all E/S0s are formed at $z>4$, predicts that `Lyman break' E/S0s will suddenly appear at the level of one per HDF-size field at $S(6.75\rm \mu m)\simeq 35\rm \mu Jy$, and therefore  that two or three may already have been detected on the Lockman Hole field of Taniguchi et al. (1998).
 However, if a high proportion of E/S0s formed from  mergers at $z<2.5$, or did not form the bulk of their stars until $z<2.5$, the numbers and/or fluxes of Lyman break E/S0s will be correspondingly reduced.


\begin{thebibliography}{}
\bibitem{}
     Acosta-Pulido J. A., et al., 1996, A\&A 315, L121.
\bibitem{}
     Aussel H., Cesarsky C. J., Elbaz D., Starck J.-L., 1998, A\&A, in press
(astro-ph/9810044).
\bibitem{}
     Bromley B. C., Press W. H., Lin H., Kirshner, R. P., 1998, ApJ, in press
(astro-ph/9711227).
\bibitem{}
     Calzetti D., Kinney A. L., Storchi-Bergmann T., 1994, ApJ, 429,582.
\bibitem{}
     Charlot S., Worthey G., Bressan A., 1996, ApJ, 457, 625.
\bibitem{}
     Elbaz D., et al. 1998.
\bibitem{}
     Flores H., Hammer F., Desert F. X., C\'esarsky C., Thuan T., Crampton D.,
 Eales S., Le F\`evre O., Omont A., Elbaz D., 1998a, A\&A, submitted
(astro-ph/9811201).
\bibitem{}
     Flores H., Hammer F., Thuan T., C\'esarsky C., Desert F. X., Omont A., 
     Lilly S. J., Eales S., Crampton D., Le F\`evre O., 1998b, ApJ, submitted
(astro-ph/9811202).
\bibitem{}
     Franceschini A., Mazzei P., de Zotti G., Danese L., 1994, ApJ, 427, 140.
\bibitem{}
     Genzel R., et al. 1998, ApJ, 498, 579.
\bibitem{}
     Glazebrook K., Abraham P., Santiago B., Ellis R., Griffiths R., 1998, 
MNRAS, 297, 885.
\bibitem{}
     Goldschmidt P., et al., 1997, MNRAS, 289, 465. 
\bibitem{}
     Infante L., de Mello D. F., Menanteau, F., 1996, ApJ, 469, L85.
\bibitem{}
     Kessler M F.., et al., 1996, A\&A 315, L27.
\bibitem{}
     Mann R.G., et al., 1997, MNRAS, 289, 482.
\bibitem{}
     Mathis J., 1990, A\&A Ann. Rev, 38, 37.
\bibitem{}
     Mazzei P., de Zotti G., 1994, ApJ, 426, 97.
\bibitem{}
     Metcalfe L., et al., 1996, A\&A, 315, L105.
\bibitem{}
     Metcalfe N., Shanks T., Fong R., Jones L., 1991, MNRAS, 249, 498.
\bibitem{}
     Metcalfe N., Shanks T., Fong R., Roche N., 1995, MNRAS, 273, 257.
\bibitem{}
     Mihos J. C., Hernquist L., 1996, ApJ, 464, 641.
\bibitem{}
     Moorwood A. F. M., 1996, A\&A 315, L109.
\bibitem{}
     Oliver S., et al., 1997, MNRAS, 289, 471.
\bibitem{}
     Page M. J., Mason K. O., McHardy I. M., Jones L. R., Carrera F.J., 
       1997, MNRAS, 291, 324.
\bibitem{}
     Pearson C., Rowan-Robinson M., 1996, MNRAS, 283, 174.
\bibitem{}
     Roche N., Eales S., 1998, MNRAS, submitted (astro-ph/9803331)
\bibitem{}
     Roche N., Eales S., Hippelein H., Willott C., 1998b, MNRAS, submitted
(astro-ph/9809271).
 \bibitem{}
     Roche N., Shanks T., Metcalfe N., Fong R., 1996, MNRAS, 280, 397.
\bibitem{}
     Roche N., Ratnatunga K., Griffiths R. E., Im M., Naim A., 1998a, MNRAS, 
293, 197.  
\bibitem{}
     Rowan-Robinson M., 1992, MNRAS, 258, 787.
\bibitem{}
     Rowan-Robinson M., Efstathiou A., 1993, MNRAS, 263, 675.
\bibitem{}
     Rowan-Robinson M., et al., 1997, MNRAS, 289, 490.
\bibitem{}
     Rush B., Malkan M. A., Spinoglio L., 1993, ApJS, 89, 1.
\bibitem{}
     Sanders D. B., Soifer B. T., Elias J. H., Madore B. F., Matthews K.,
 Neugebauer G., Scoville N. Z., 1988, ApJ, 325, 74.
\bibitem{}
     Sanders D. B., Mirabel I. F., 1996, A\&A Ann Rev, 34, 749.
\bibitem{}
     Schade D., Lilly S. J., Le F\`evre O., Hammer F., Crampton D., 1995,
ApJ, 451, L1.
\bibitem{}
     Schade D., Carlberg, R. G., Yee, H. K. C., L\'{o}pez-Cruz, O., 1996, 
ApJ, 464, L63.
\bibitem{}
     Schmitt H. R., Kinney A. L., Calzetti D., Storchi-Bergmann T., 1997, AJ,
114, 592. (SKCS)
\bibitem{}
     Spinoglio L., Mlkan M., Ruch B., Carrasco L., Recillas-Cruz E., 1995,
ApJ, 453, 616.
\bibitem{}
     Taniguchi Y., et al., 1997, A\&A, 328, L9. 
\bibitem{}
     Weedman D. W., Wolovitz J., Bershady M., Schneider D., 1998, AJ,
 in press (astro-ph/9806398).
\bibitem{}
     Xu C., Hacking P., Fang F., Shupe D., Lonsdale C., Lu N., Helou G.,
Stacey G., Ashby M., 1998, ApJ, in press (astro-ph/9806194).
\end{thebibliography}
\end{document}